\documentclass[a4paper]{jpconf}
\usepackage{graphicx}
\usepackage[square,sort&compress]{natbib}

\begin{document}
\title{Visualization and spectral synthesis of rotationally distorted stars}

\author{T. H. Dall$^1$, L. Sbordone$^{2,3}$}

\address{$^1$ESO, Karl-Schwarzschild-Str. 2, 85748 Garching bei M\"unchen, Germany}
\address{$^2$Max Planck Institute for Astrophysics, Karl-Schwarzschild-Str. 1, 85741 Garching bei M\"unchen, Germany}
\address{$^3$ZAH - Landessternwarte Heidelberg, K\"onigstuhl 12, 69117 Heidelberg, Germany}

\ead{tdall@eso.org}

\begin{abstract}
Simple spherical, non-rotating stellar models are
inadequate when describing real stars in the limit of very fast
rotation: Both the observable spectrum and the geometrical shape of
the star deviate strongly from simple models. We attempt to approach
the problem of modeling geometrically distorted, rapidly rotating
stars from a new angle: By constructing distorted geometrical models
and integrating standard stellar models with varying temperature,
gravity, and abundances, over the entire surface, we attempt a
semi-empirical approach to modeling. Here we present our methodology, and present simple examples of applications.
\end{abstract}

\section{Introduction}
For many years, stellar atmospheric models and spectrum synthesis has been synonymous with the use of standard 1D atmospheric models. Lately,  3D models have added another dimension of realism to this picture. However, all models in use today are local models, treating only  a small part of the stellar surface. Calculation of synthetic spectra then relies on assuming that the stellar atmosphere looks  the same everywhere.  Real stellar atmospheres are of course not  the same everywhere. 
In this contribution we wish to highlight an aspect of stellar spectral synthesis that is mostly overlooked, namely the deviations from spherical symmetry.

The problems associated with stars that are not spherically symmetric were first addressed by von Zeipel \cite{vonZeipel1,vonzeipel2}, who showed that stars that were flattened due to fast rotation would have a latitude dependent brightness distribution.  However, the observed spectrum of a star will be influenced by many different additional effects from differential rotation, turbulence, temperature and gravity variations --- all latitude dependent. Not to mention  effects like spots, plagues, etc. that are both latitude and longitude dependent.


\section{Methodology}
Ideally, one should be able to take an observed spectrum and perform an inverse analysis, that is, from the spectrum infer the exact distributions of temparature, gravity, abundances, etc., as well as the shape of the star.  However, such capabilities are far into the future since we still don't understand fully all the effects going into the structure of a star and the formation of its spectrum.  

With this work, we are attempting a different approach to learn more about these effects, that determine the structure of geometrically distorted stars and the possibilities of observing those effects. Our approach is:
\begin{itemize}
\item We assume a shape of the star -- any shape.
\item We assume (any) distributions of temperature, gravity, abudances, etc. across the surface.
\item We assume any rotation law on the form $v(\theta)$.
\item At any point on the stellar surface we then compute a resulting intensity spectrum, in the direction of the observer. For this, any kind of local stellar model atmosphere can be used, whether ATLAS, MARCS, or state-of-the-art 3D models. 
\item We then sum up all the contributions from the visible part of the star into an "observed" synthetic spectrum.
\end{itemize}

This is an \emph{exploratory} approach, aimed at providing more insight into the individual effects involed in shaping the spectra of real stars, rather that an attempt at producing realistic models of real stars, although that is of course the long term aim.


\begin{figure}
\begin{center}
\includegraphics[width=10cm]{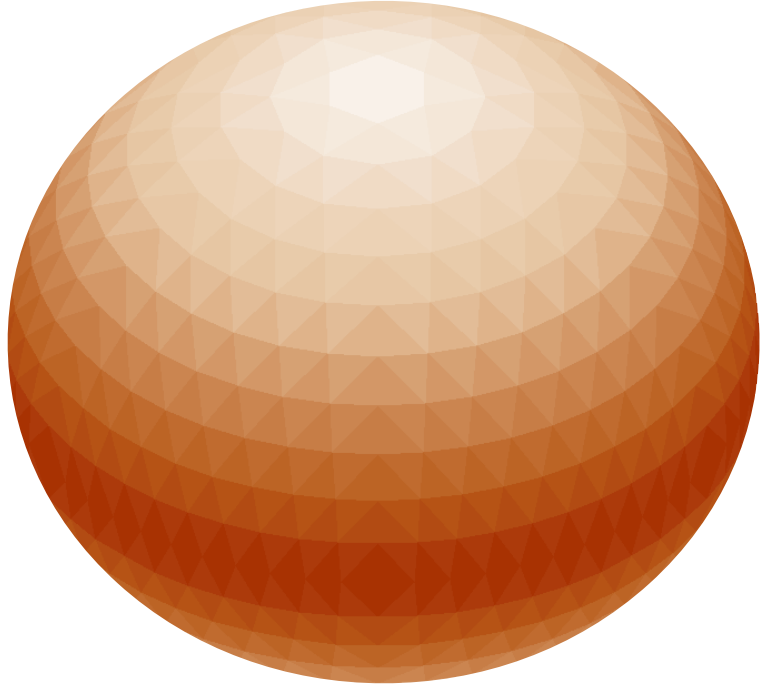}
\end{center}
\caption{\label{fig:tiles}Example of a tiled spherical surface with 22 latitudinal bands ($N=11$) before the flattening coordinate transformation is applied. Note that the shape of the triangular tiles varies slightly across the surface.}
\end{figure}

We want to avoid dependence on analytical descriptions of the parameters, which means that we must be able to freely assign any value of any parameter to any  point on the surface. For this reason we chose to create a tiled surface, i.e., we cover the entire surface with plane segments to which we can assign any combination of parameters we wish. In principle the only limit to the fine graininess of the surface is the  speed of the computer hardware.  Since a sphere cannot be tiled with equal-area regular polygons, we chose instead to employ triangular tiling where the individual tiles have slightly different surface areas. The tiles are arranged in latitudinal bands, where $N$ is the number of bands per hemisphere. See Fig.~\ref{fig:tiles} for an example. 

\begin{figure}
\begin{center}
\includegraphics[width=16cm]{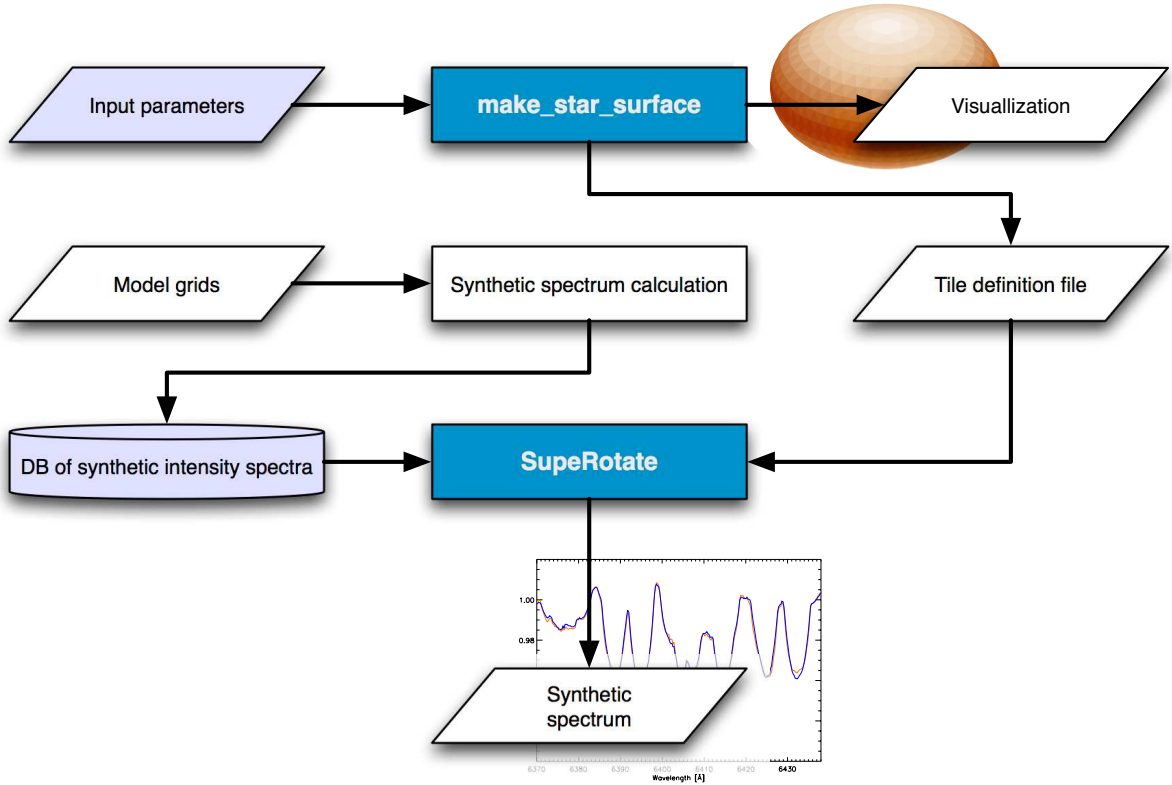}
\end{center}
\caption{\label{fig:flowch}Flow chart describing the process. The blue boxes represent the software modules described in this paper. The box SupeRotate is where the actual spectral synthesis is done. The database of synthetic intensity spectra can be constructed using any current model atmosphere grid.}
\end{figure}

The full procedure of what we have named ``SupeRotate'' is shown schematically in Fig.~\ref{fig:flowch}. The details will be treated in a later paper.

\section{Results}
As a zero-order test of the code we first simulated a non-flattened star with no variation of parameters across the surface and compared this to a standard synthetic spectrum derived from the same set of models. We tested this for models with varying rotational velocities, assuming rigid rotation. The results were practically identical for $N \ge 9$. 

As more interesting tests we have simulated two cases which resemble well known stars, namely Altair and FK Com. However, these models are intended to test the model sanity rather than produce actual physical results on the two stars.

\subsection{Models of "Altair"}
Altair is an extensively studied flattened, fast rotating early type star. Its oblateness or flattening of $r_\mathrm{pol}/r_\mathrm{eq} = 0.80$ was first demonstrated by van~Belle et al.~\cite{altair2001} with subsequent studies confirming these findings \cite{altair2006,altair2007}.
For this study, for the purpose of illustration, we are modeling Altair with an equatorial rotational velocity of $80$~km\,s$^{-1}$ as opposed to the true velocity of about $270$~km\,s$^{-1}$. Thus, we refer to our models as "Altair" rather than Altair.
\begin{figure}
\begin{center}
\includegraphics[width=16cm]{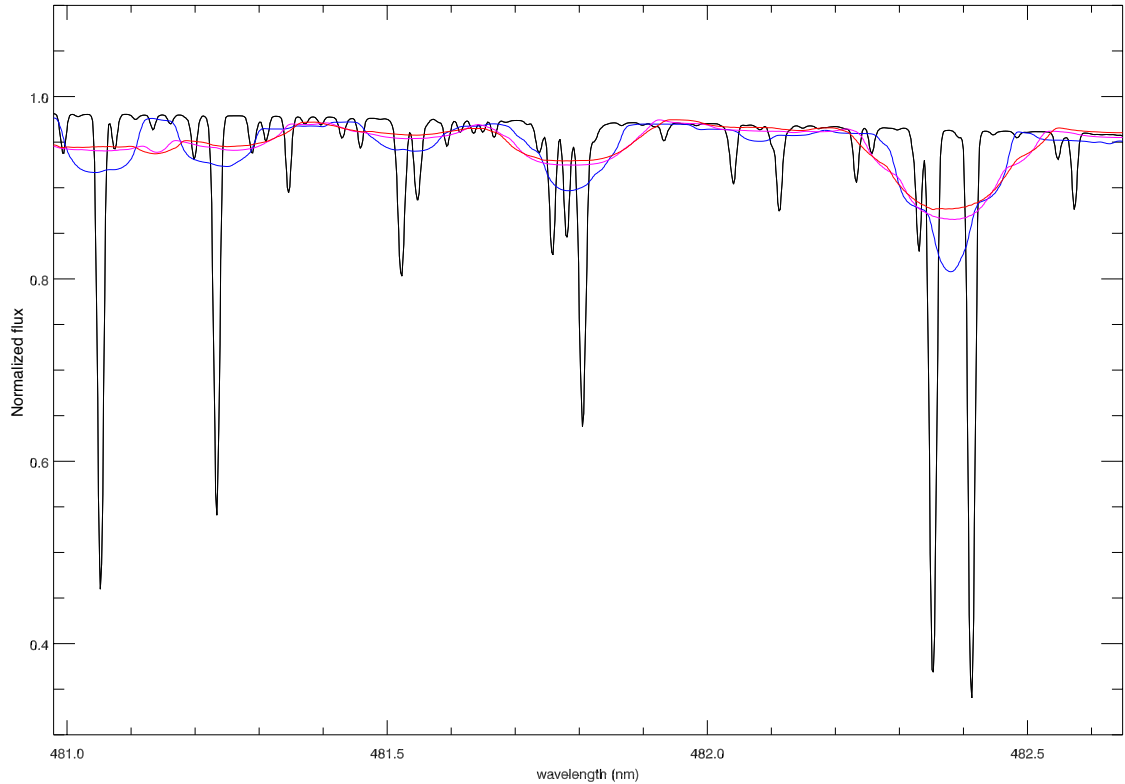}
\end{center}
\caption{\label{fig:alt_closeup}Close up of a section of the resulting synthetic spectrum of "Altair" with $T_\mathrm{eff}$ and $\log g$ varying from pole to equator, and at four different values of the inclination; $i = 0^\circ$ (black), $i = 30^\circ$ (blue), $i = 60^\circ$ (magenta), $i = 90^\circ$ (red).}
\end{figure}

We have calculated synthetic spectra for four different inclinations (see Fig.~\ref{fig:alt_closeup}) with atmospheric parameters varying linearly in latitude from pole to equator. The parameters used are summarized in Table~\ref{tab:altair}.
\begin{table}
\caption{\label{tab:altair}Parameters used in the simulations of "Altair". The model is then used to calculate synthetic spectra for four values of the inclination; $i = 0^\circ, 30^\circ, 60^\circ, 90^\circ$.}
\begin{center}
\begin{tabular}{ll}
\br
Parameter & Range (pole --- equator) \\ 
\mr
$T_\mathrm{eff}$ & 7300 K --- 6900 K \\
$\log g$ & 4.1 --- 3.9 \\
\br
\end{tabular}
\end{center}
\end{table}

Further tests on "Altair" is planned and will be reported elsewhere.

\subsection{Models of "FK Com"}
FK Com is, with its $v \sin i = 160$~km\,s$^{-1}$, the fastest rotating and most active of the FK Comae stars (see \cite{fkcom2009} and references therein). These are a small group of magnetically very active late-type giants with rotation periods of only a few days. The photometric and spectroscopic characteristics of FK Comae stars are very similar to those of the very active RS CVn stars, but the RS CVn stars are close binary systems while FK Comae stars are single, possibly the result of a merging of the components of a RS CVn system.

\begin{figure}
\begin{center}
\includegraphics[width=14cm]{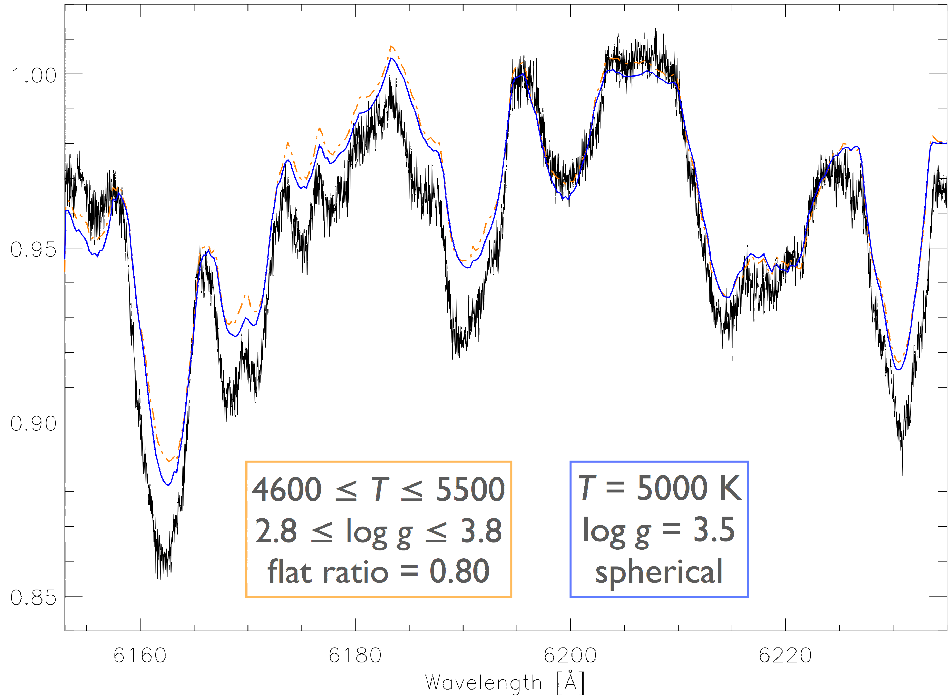}
\end{center}
\caption{\label{fig:fkcom_obs}Observed spectrum of FK Com (black) along with our synthetic spectrum from SupeRotate (orange) and from standard synthesis (blue).}
\end{figure}

The parameters of FK Com is well known and lies within the model grid we have prepared for our testing phase. Though it is believed that temperature, gravity and indeed abundances are varying widely over the surface, we have employed parameters varying smoothly with $\theta$ only. Thus, our model is "FK Com" rather than FK Com.  Table~\ref{tab:fkcom} lists the parameters used, as well as the standard average stellar model atmosphere consensus for FK Com.
\begin{table}
\caption{\label{tab:fkcom}Parameters used in the simulations of "FK Com" as well as the average parameters from \cite{fkcom2007}. The spectra are all calculated for the known inclination $i = 60^\circ$.}
\begin{center}
\begin{tabular}{ll|l}
\br
Parameter & Range (pole --- equator) & Standard model \\ 
\mr
$T_\mathrm{eff}$ & 5500 K --- 4600 K & 5000 K \\
$\log g$ & 3.8 --- 2.8 & 3.5 \\
\br
\end{tabular}
\end{center}
\end{table}
In our calculation we used the known $v \sin i$ and inclination $i = 60^\circ$ as well as an assumed flattening of $0.80$. The resemblance with the observed spectrum is good as is evident from Fig.~\ref{fig:fkcom_obs}.

\begin{figure}
\begin{center}
\includegraphics[width=14cm]{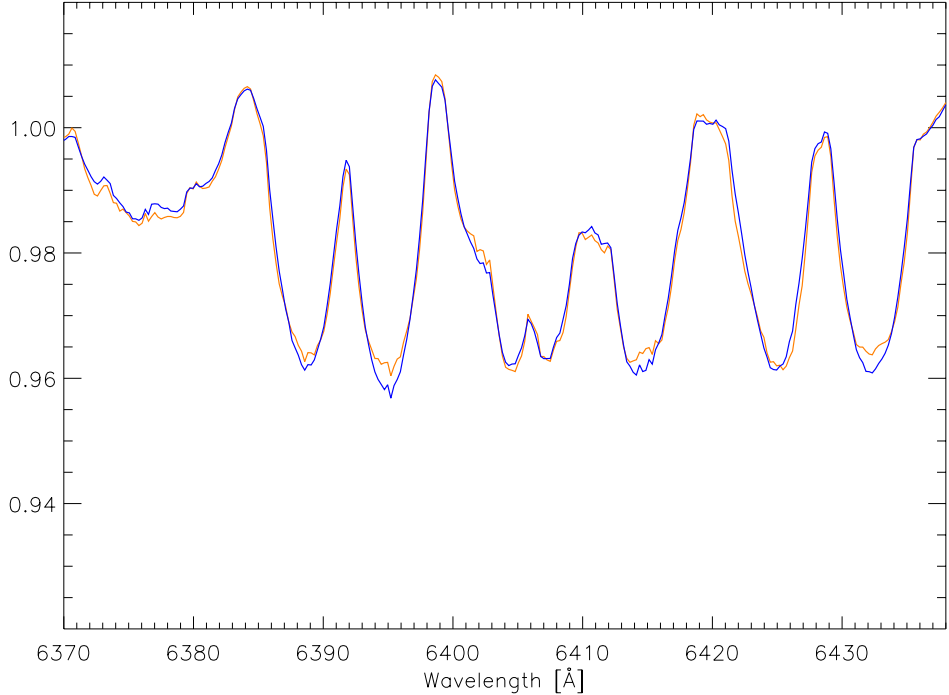}
\end{center}
\caption{\label{fig:fkcom_mod}Synthetic spectrum from SupeRotate (orange) and from standard synthesis (blue) in a 70~\AA\ interval.}
\end{figure}

In Fig.~\ref{fig:fkcom_mod} we show the results of the flattened model campared with the standard model for a small spectral range. It is clear that the differences are quite small and possibly too small to be observed, in particular when one considers that the surface of FK Com is a mix of different temperatures, abundance patches, active regions, etc.  However, what this code can provide is firm ideas about the relative significance of all these effects and what changes might be induced by variations of any of these parameters.

\section{Conclusion}
In this contribution we have explored a new modeling approach which is aimed primarily at gaining insight into the underlying individual effects that are shaping the spectra of stars with non-spherical and non-homogenious surfaces. 

Our models are currently at the "toy model" stage, suited for an exploratory approach. Our models are easy to use, flexible, and highly adaptable. Following further testing, the software will be made freely available to the community. Meanwhile, we welcome requests and suggestions for suitable projects which we will undertake as extended testing.

\ack
We wish to thank H.~Korhonen for kind permission to use the spectrum of FK Com.

\bibliography{ref}


\end{document}